\documentclass[twocolumn, nofootinbib, secnumarabic, amssymb, preprintnumbers, superscriptaddress, aps, prd]{revtex4-1}

\usepackage{graphicx}

\usepackage{cancel,tabularx,moreverb,fancybox,amsmath,float,bm,braket,slashbox,txfonts,amssymb,bm,accents}
\usepackage{latexsym}
\usepackage{here}

\newcommand{\ex}[1]{\mathrm{e}^{#1}}

\newcommand{\pa}[1]{\left(#1 \right)}

\newcommand{\br}[1]{\left[#1 \right]}
\newcommand{\BR}[1]{\Biggl[#1 \Biggr]}

\newcommand{\bb}[1]{\mathbb{#1}}

\newcommand{\ca}[1]{\mathcal{#1}}

\newcommand{\abs}[1]{\left|#1\right|}
\newcommand{\ave}[1]{\langle #1\rangle}
\newcommand{\ar}[1]{\xrightarrow[#1]{}}

\newcommand{\dg}[1]{#1^{\dagger}}
\newcommand{\fr}{\frac}
\newcommand{\s}[1]{\sqrt{#1}}

\def\be{\begin{equation}}
\def\ee{\end{equation}}
\def\ba{\begin{eqnarray}}
\def\ea{\end{eqnarray}}

 \def\ep{{\epsilon}}

 \def\a{{\alpha}}

 \def\b{{\beta}}
 \def\e{{\epsilon}}

\def\tr{{\text{tr}}}

\def\dd{{\mathrm{d}}}

\begin{document}

\preprint{YITP-19-71}
\preprint{UT-19-19}
\title{Entanglement Entropy after Double-Excitation as Interaction Measure}
\date{\today}
\author{Yuya Kusuki}\email[]{yuya.kusuki@yukawa.kyoto-u.ac.jp}
\affiliation{\it
Center for Gravitational Physics, \\
Yukawa Institute for Theoretical Physics (YITP), Kyoto University, \\
Kitashirakawa Oiwakecho, Sakyo-ku, Kyoto 606-8502, Japan.
}
\author{Masamichi Miyaji}\email[]{masamichi.miyaji@hep-th.phys.s.u-tokyo.ac.jp}
\affiliation{\it
Department of Physics, Faculty of Science,
University of Tokyo, Tokyo 113-0033, Japan}

\begin{abstract}
We study entanglement entropy after a double local quench in two-dimensional conformal field theories (CFTs), with any central charge $c>1$. 
In the holographic CFT, such a state with double-excitation is dual to an AdS space with two massive particles introduced from the boundary.
We show that the growth after the double local excitations cannot be given by the sum of two local quenches but with an additional {\it negative} term.
This negative contribution can be naturally interpreted as due to the attractive force of gravity.
In CFT side, this evaluation of the entanglement entropy is accomplished by a special limit of 6-point functions, where we employed the fusion matrix approach for multi-point conformal blocks developed in arXiv:1905.02191.
\end{abstract}

\maketitle
\noindent

\section{Introduction}
The AdS/CFT correspondence (holographic principle) \cite{Maldacena1999b} has provided new insights in quantum gravity, since we understand CFTs relatively
well.
It is expected that there are holographic duals of quantum gravity theories in AdS${}_3$ (i.e., holographic CFTs), and in general, they are expected to be unitary, compact CFTs with central charge $c>1$ and without chiral primaries (hence, no extra currents apart from the Virasoro current), which we call {\it pure CFTs} (for e.g., \cite{Collier2016,CollierKravchukLinYin2017,Kusuki2019b}) .

One interesting question is how the dynamics of a two-particle state in AdS is realized on the CFT side.
In particular, we make an effort to understand the dynamics in AdS by making use of the entanglement entropy, which is one very useful tool to access the global structure of a given quantum field theory \cite{Ryu2006, Hubeny2007}.
The entanglement entropy is defined by
\begin{equation}
S_A=-\tr \rho_A \log \rho_A,
\end{equation}
where $\rho_A$ is a reduced density matrix for a subsystem $A$, obtained by tracing out its complement. 
In practice, this quantity is calculated as the $n \to 1$ limit of the Renyi entropy, which is a generalization of the entanglement entropy as
\begin{equation}
S_A^{(n)}=\fr{1}{1-n} \log \tr \rho_A^n.
\end{equation}
We particularly focus on the dynamics of the entanglement, therefore, we consider a locally excited state $\ket{\Psi}$, which is defined by acting with a local operator $ O(-l)$ on the CFT vacuum $\ket{0}$ in the following manner,
\be\label{lopw}
\ket{\Psi(t)}=\s{\ca{N}}\ex{-\ep H-iHt} O(-l)\ket{0}, 
\ee
where $\ca{N}$ is the normalization factor and an infinitesimal positive parameter $\e$ is an ultraviolet regularization. We choose the subsystem $A$ to be the half-space $x>0$ and apply the excitation in its complement.
The growth of the entanglement attributed to the excitation can be seen from the difference between the excited state and the ground state,
 \be\label{eq:difs}
 \Delta S^{(n)}_A[O](t)=S^{(n)}_A(\ket{\Psi(t)})-S^{(n)}_A(\ket{0}),
 \ee
which is our main interest in this paper.

In CFT, this quantity can be calculated in terms of a correlator \cite{Nozaki2014a,Nozaki2014} and has already studied in RCFTs \cite{He2014,Numasawa2016}, holographic CFTs \cite{Caputa2014a, Asplund2015}, and also non-perturbatively in pure CFTs \cite{Kusuki2018c,Kusuki2018b, Kusuki2019b}. (See also  \cite{Caputa2014a, David2016, Caputa2017, He2017, Guo2018, Shimaji2018, Apolo2018}, which revealed the growth of the entanglement entropy after a local quench in other setups.)
It is interesting to note that the entanglement growth in RCFTs approaches constant in the late time limit, on the other hand, the entanglement entropy in holographic CFTs shows a logarithmic growth. This difference is thought to be due to the chaotic nature of holographic CFTs.  In other words, we expect that this late time behavior can also be used as a criterion of chaotic nature of a given quantum field theory. 

One may wonder what happens if we consider a locally double-excited state, instead of a single-excited state,
\be\label{eq:Dstate}
\ket{\Psi(t)}=\s{\ca{N}}\ex{-\ep H-iHt} O_A(-l_A) \ex{\ep H + iHt}  \ex{-\ep H-iHt}  O_B(-l_B) \ket{0}.
\ee
At least in RCFTs, it is naturally expected that the entanglement growth for this state can be given by just a  simple sum rule as
\begin{equation}
\Delta S_A[O_A;O_B]=\Delta S_A[O_A] + \Delta S_A[O_B],
\end{equation}
where $\Delta S_A$ is the growth of the entanglement entropy after a double-excitation with $O_A$ and $O_B$.
This is because the entanglement entropy in RCFTs behaves as if correlations were carried by free quasiparticles \cite{Asplund2015a}.
Actually this rule has already shown in \cite{Caputa2016a, Numasawa2016, Guo2018}.

However, there is no reason that the sum rule can be also applied to other CFT, in particular, holographic CFTs because the quasiparticle picture breaks down in holographic CFTs. For this reason, it would rather be natural that the sum rule does not work in holographic CFTs. The gravity dual of the excited state (\ref{lopw}) is obtained by a heavy falling particle in AdS \cite{Caputa2014a}, therefore, the double-excited state (\ref{eq:Dstate}) should correspond to two falling particles. Since these two particles interact with each other by the gravity force, we expect that the effect of this interaction would also appear in the entanglement growth $\Delta S_A[O_A;O_B]$, that is, it is not given by just the sum rule.

On this background, the main aim of this paper is to verify whether the entanglement entropy measures the interaction between two excitations.
This question has remained unexplored because we have little knowledge of the Regge singularity of a correlator, which is necessary to calculate the Renyi entropy.
Recently, we investigated this singularity numerically in \cite{Kusuki2018,Kusuki2018b} and give the analytic form in \cite{Kusuki2019b} from our new findings of many properties of the conformal blocks \cite{Kusuki2018a, Kusuki2018c} (see also \cite{Collier2018}).
The key point is that the singularity of a correlator in the Regge limit is controlled by the pole structure of the { \it monodromy matrix}. By utilizing this development, we accomplish to provide the entanglement entropy for a double-excited state.

\section{Single-Excitation}

\begin{figure}[t]
 \begin{center}
  \includegraphics[width=5.0cm]{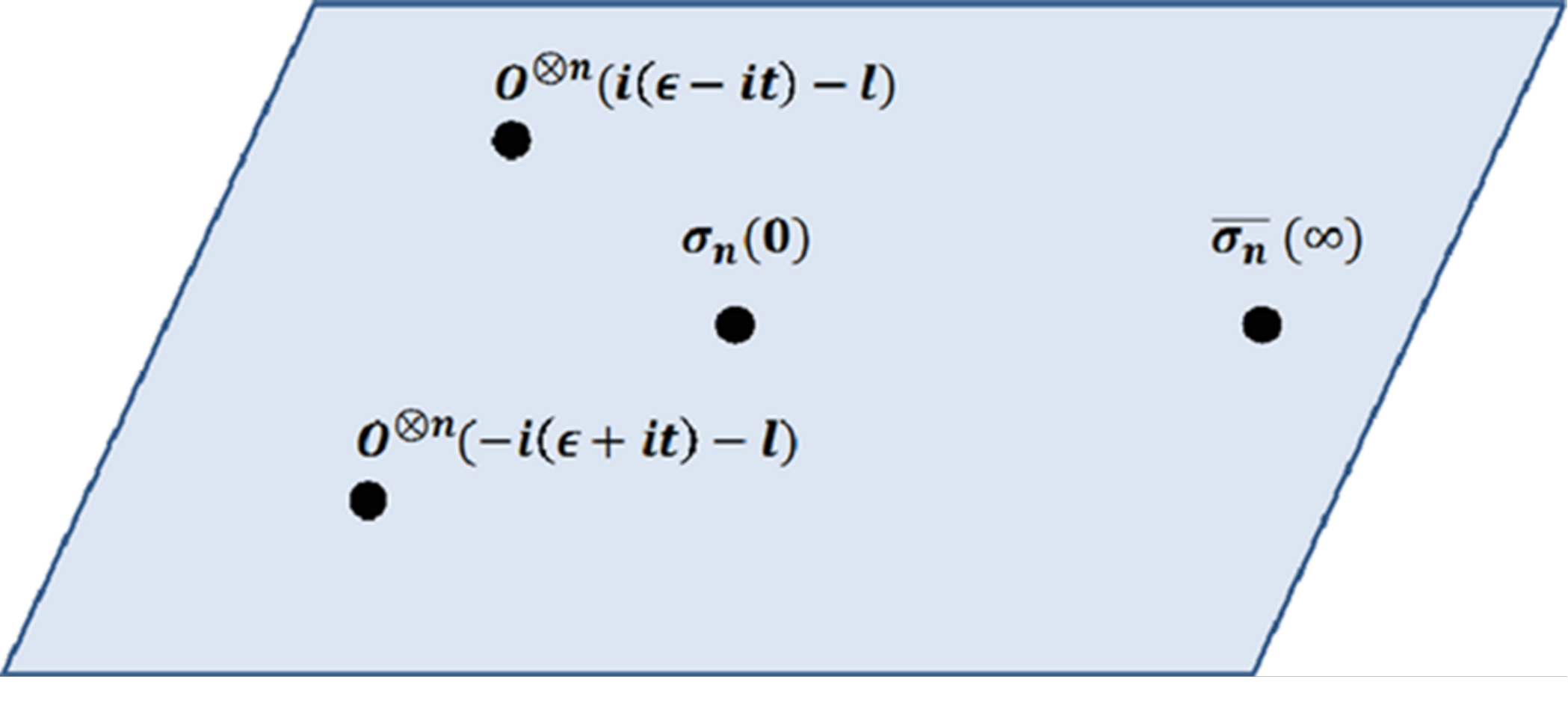}
 \end{center}
 \caption{The holomorphic part of the positions of operators in the replica computation (\ref{eq:defREE}).}
 \label{fig:pos}
\end{figure}

In this section, we will give a brief review of the calculations of the Renyi entropy after a single-excitation. 
Here, we consider a local excitation $O$ separated by a distance $l$ from the boundary of $A$, as shown in the left of Figure \ref{fig:pos}.
The Renyi entropy in this setup can be calculated using twist operators as \cite{Asplund2015, Kusuki2018b, Kusuki2019b}
\begin{equation}\label{eq:defREE}
\Delta S^{(n)}_A[O]=\frac{1}{1-n}\log \frac{\ave{\bb{O}\bb{O}\sigma_n \bar{\sigma_n} }}{\ave{\bb{O}\bb{O}}\ave{\sigma_n \bar{\sigma_n}}},
\end{equation}
where we introduce some notations on the cyclic orbifold CFT $\ca{M}^n/\bb{Z}_n$ in TABLE \ref{table:orbifold},
\begin{table}[h]
  \begin{tabular}{|l|l|} 
\hline
     $\ca{F}^{(n)}$ & \parbox{7.8cm}{The conformal block associated to the chiral algebra $(\text{Vir})^n/\bb{Z}_n$, instead of the Virasoro.  (Note that this conformal block is defined on the CFT with the central charge $nc$.)	}	\\  \hline
    ${\bold F}^{(n)}$ &  \parbox{7.8cm}{The fusion matrix associated to the chiral algebra $(\text{Vir})^n/\bb{Z}_n$.} \\ \hline
    ${\bold M}^{(n)}$ &  \parbox{7.8cm}{The monodromy matrix associated to the chiral algebra $(\text{Vir})^n/\bb{Z}_n$.} \\ \hline
    $h_{\sigma_n}$ &  \parbox{7.8cm}{The conformal dim. of the twist op.; $h_{\sigma_n}\equiv \frac{c}{24}\pa{1-\fr{1}{n^2}}$.} \\ \hline
    $\a_n$ &  \parbox{7.8cm}{The Liouville momentum of $h_{\sigma_n}$;  $h_{\sigma_n}\equiv \a_n (Q-\a_n)$.  } \\ \hline
    $\bb{O} $ &  \parbox{7.8cm}{ The operator defined on the cyclic orbifold CFT $\ca{M}^n/\bb{Z}_n$, using the operators in the seed CFT $\ca{M}$ as $O^{\otimes n}= O \otimes O \otimes \cdots \otimes O (\equiv \bb{O})$.
} \\ \hline
  \end{tabular}
\caption{The notations on the cyclic orbifold CFT $\ca{M}^n/\bb{Z}_n$. We will use these notations in the following.}
\label{table:orbifold}
\end{table}
\\
And we also introduce the notations usually found in Liouville CFTs,
\begin{equation}
c=1+6Q^2, \ \ \ \ \ Q=b+\fr{1}{b}, \ \ \ \ \ h_i=\a_i(Q-\a_i).
\end{equation}

By using the cross ratio $z=\frac{z_{12}z_{34}}{z_{13}z_{24}}$, we can rewrite (\ref{eq:defREE}) as
\begin{equation}\label{eq:GRenyi}
\frac{\ave{\bb{O}\bb{O}\sigma_n \bar{\sigma_n} }}{\ave{\bb{O}\bb{O}}\ave{\sigma_n \bar{\sigma_n}}}=\abs{z^{2h_{\sigma_n}}}^2G(z,\bar{z}),
\end{equation}
where $G(z,\bar{z})$ is the four point function
\be
G(z,\bar{z})=\braket{\bb{O}(\infty) \bb{O}(1)\bar{\sigma}_n(z,\bar{z})\sigma_n(0)},
\ee
and in our setup, the cross ratio $(z,\bar{z})$ is given by
\begin{equation}
z=\frac{2i \e}{l-t+i \e}, ~~~~~\bar{z}=-\frac{2i\e}{l+t-i\e}.
\end{equation}
From these expressions, one finds that the sign of the imaginary part of the cross ratio $z$ changes at $t=l$.  As a result, the holomorphic cross ratio picks up the factor $\ex{-2\pi i}$ at $t=l$ as $1-z \to \ex{-2\pi i}(1-z)$. On the other hand, this does not happen for the anti-chiral coordinate $\bar{z}$. Note that the limit $z \to 0$ after picking up this holomorphic monodromy is usually called the {\it Regge limit}.

In general, the four point function $G(z,\bar{z})$ can be decomposed into conformal blocks as
\begin{equation}
\begin{aligned}
\sum_\b \ C_{\b \sigma_n \bar{\sigma}_n} C_{\b \bb{O} \bb{O}^{\dagger}}
 {\ca{F}^{(n)}}^{\bar{\sigma}_n \sigma_n }_{\bb{O}\dg{\bb{O}}}(h_\b|z)  \overline{{\ca{F}^{(n)}}^{\bar{\sigma}_n \sigma_n }_{\bb{O}\dg{\bb{O}}}} (h_\b|\bar{z}),
\end{aligned}
\end{equation}
where $\b$ runs over all  $(\text{Vir})^n/\bb{Z}_n$ primary operators.
The monodromy effect to the conformal block is expressed in terms of the monodromy matrix as \cite{Kusuki2019b}
\newsavebox{\boxpd}
\sbox{\boxpd}{\includegraphics[width=90pt]{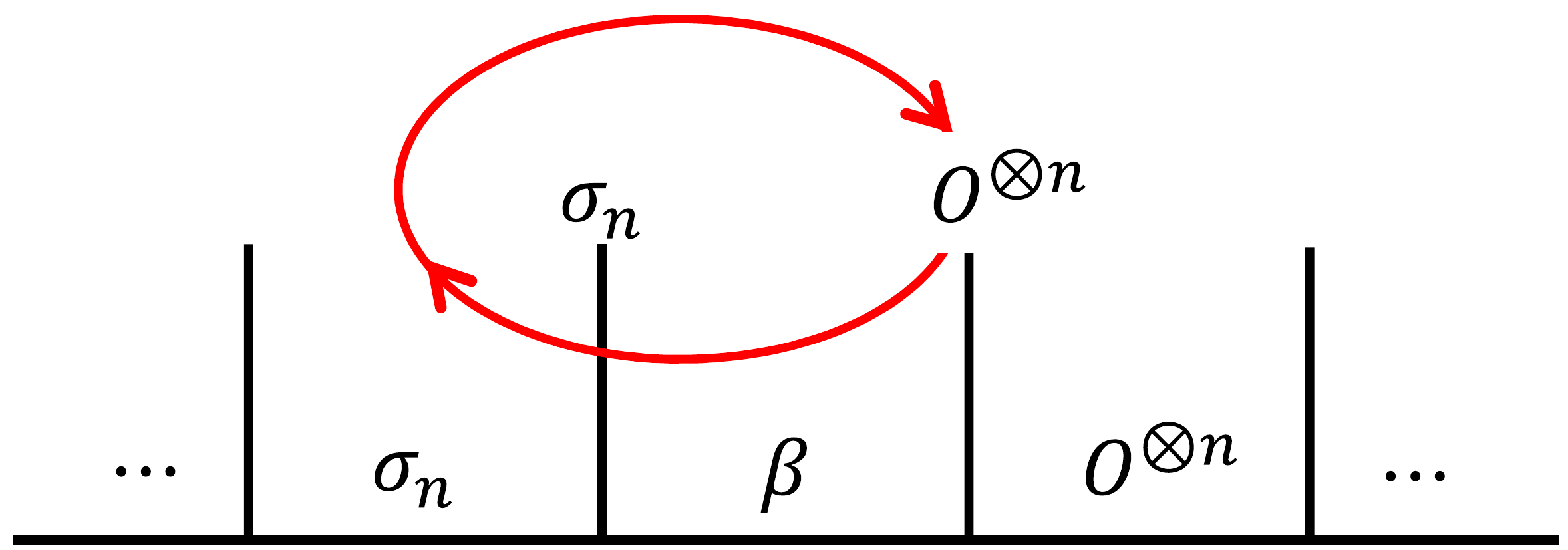}}
\newlength{\pdw}
\settowidth{\pdw}{\usebox{\boxpd}} 

\newsavebox{\boxpe}
\sbox{\boxpe}{\includegraphics[width=90pt]{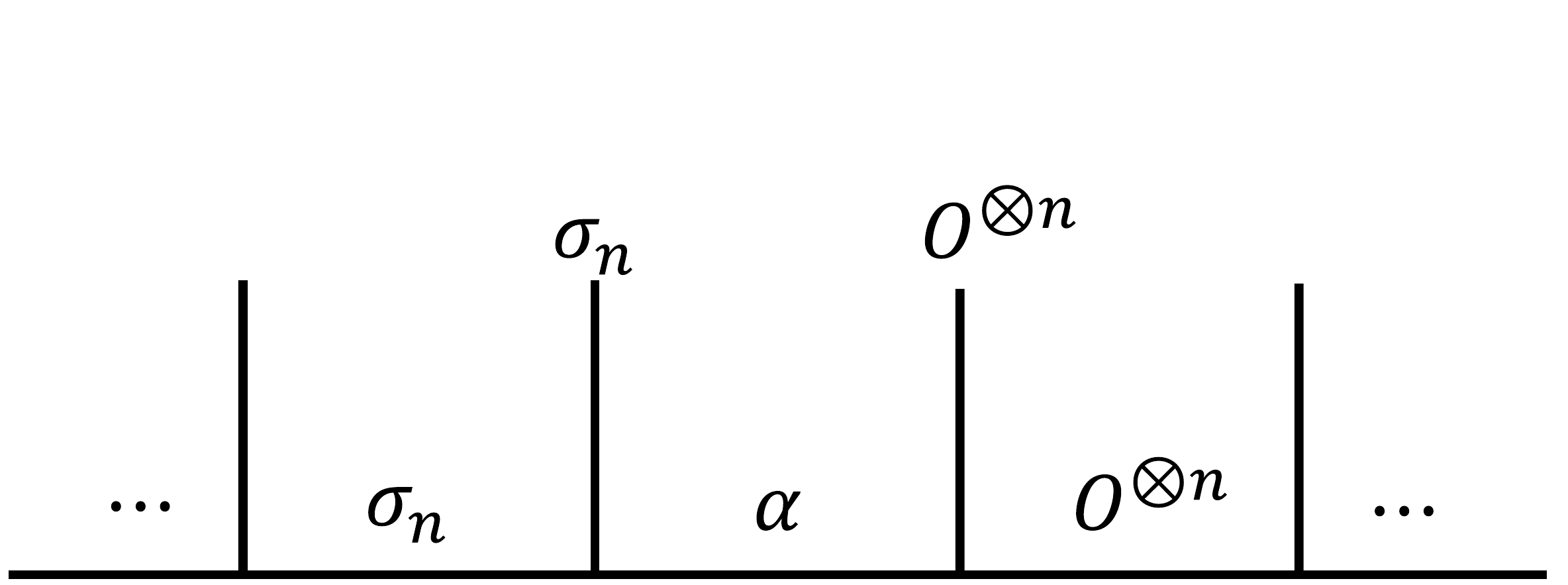}}
\newlength{\pew}
\settowidth{\pew}{\usebox{\boxpe}} 

\begin{equation}\label{eq:monotrans}
\begin{aligned}
&\parbox{\pdw}{\usebox{\boxpd}} \\
&= \int \dd \a \  {{\bold M}^{(n)}}_{\b, \a}[O]  \parbox{\pew}{\usebox{\boxpe}}.
\end{aligned}
\end{equation}
The key point is that if we restrict ourselves to pure CFTs, the minimal value of $\a$ is realized by a particular residue at $\a=\a^{\text{min}}$,
\begin{equation}
\a^{\text{min}}= \min\{2\a_O, 2\a_n   \}.
\end{equation}

From now on, we will focus on the late time region ($t>l$).
In the $\e \to 0$ limit, the four point function can be approximated by the vacuum contribution as
\begin{equation}\label{eq:twistRenyi}
\begin{aligned}
\int \dd \a \ {{\bold M}^{(n)}}_{0, \a}[O]
 {\ca{F}^{(n)}}^{\bar{\sigma}_n \sigma_n }_{\bb{O}\dg{\bb{O}}}(n h_{\a}|z)  \overline{{\ca{F}^{(n)}}^{\bar{\sigma}_n \sigma_n }_{\bb{O}\dg{\bb{O}}}} (0|\bar{z}).
\end{aligned}
\end{equation}
Here, we used the following asymptotics of the conformal block $\ca{F}^{(n)}$,
\begin{equation}
 {\ca{F}^{(n)}}^{\bar{\sigma}_n \sigma_n }_{\bb{O}\dg{\bb{O}}}(n h_{\a}|z)
\ar{z \to 0}
z^{nh_\a - 2h_{\sigma_n}}.
\end{equation}
This also means that in the limit $\e \to 0$, the dominant contribution of the integral over $\a$ comes from the minimal one $\a=\min\{2\a_O, 2\a_n   \}$.
As a result, the entanglement entropy (the $n \to1$ limit of the Renyi entropy) is given by
\begin{equation}\label{eq:EEinpureCFT}
\begin{aligned}
\fr{c}{6}\log\fr{ti}{2 \e}+\lim_{n \to 1} \fr{1}{1-n} \log  \BR{  \text{Res}\pa{  -2\pi i \   {{\bold M}^{(n)}}_{0,\a}[O];2\a_n }}.
\end{aligned}
\end{equation}
And also we can give the $n$-th ($n\geq2$) Renyi entropy after a local quench $h_O<\fr{c-1}{32}$ as
\begin{equation}\label{eq:REEheavy}
\begin{aligned}
\fr{n h_{2\a_O}}{1-n}\log  \pa{\fr{2 \e}{t i}}    + \fr{1 }{1-n}\log \BR{ \text{Res}\pa{  -2\pi i \   {{\bold M}^{(n)}}_{0,\a}[O]; 2\a_{O}}}.
\end{aligned}
\end{equation}
These results are consistent with the results from the light cone limit approach and the gravity dual calculation \cite{Kusuki2019b}.
Moreover, we can find that in the classical limit ($c \to \infty$), the monodromy matrix associated to $(\text{Vir})^n/\bb{Z}_n$ can be expressed in terms of the Bekenstein-Hawking entropy as
\begin{equation}\lim_{n \to 1} \fr{1}{1-n} \log  \BR{ -2i \text{Res}\pa{  -2\pi i \   {{\bold M}^{(n)}}_{0,\a}[O];2\a_n }}\ar{\substack{c \to \infty \\ h_O \gg c } } S_{\text{BH}}\pa{O},
\end{equation}
where $S_{\text{BH}}\pa{O} $ is given by the Cardy formula, 
\begin{equation}
S_{\text{BH}}\pa{O}=2\pi\s{\fr{c}{6}\pa{h_{ O}-\fr{c}{24}}}.
\end{equation}

\section{Double-Excitation}
Here, we consider a double excited state with two operators $O_A(-l_A)$ and $O_B(-l_B)$ ($l_A<l_B$).
The entanglement entropy for this state can be calculated by the following six-point correlator,
\begin{equation}
\Delta S^{(n)}_A[O_A;O_B]=\frac{1}{1-n}\log \frac{\ave{O_A^{\otimes n}O_A^{\otimes n} O_B^{\otimes n}O_B^{\otimes n}   \sigma_n \bar{\sigma_n} }}{\ave{O_A^{\otimes n}O_A^{\otimes n} O_B^{\otimes n}O_B^{\otimes n} }\ave{\sigma_n \bar{\sigma_n}}}.
\end{equation}
In the late time limit (i.e., the Regge limit), this correlator can be also evaluated in the same way as for the single excitation case;

\newsavebox{\boxpa}
\sbox{\boxpa}{\includegraphics[width=100pt]{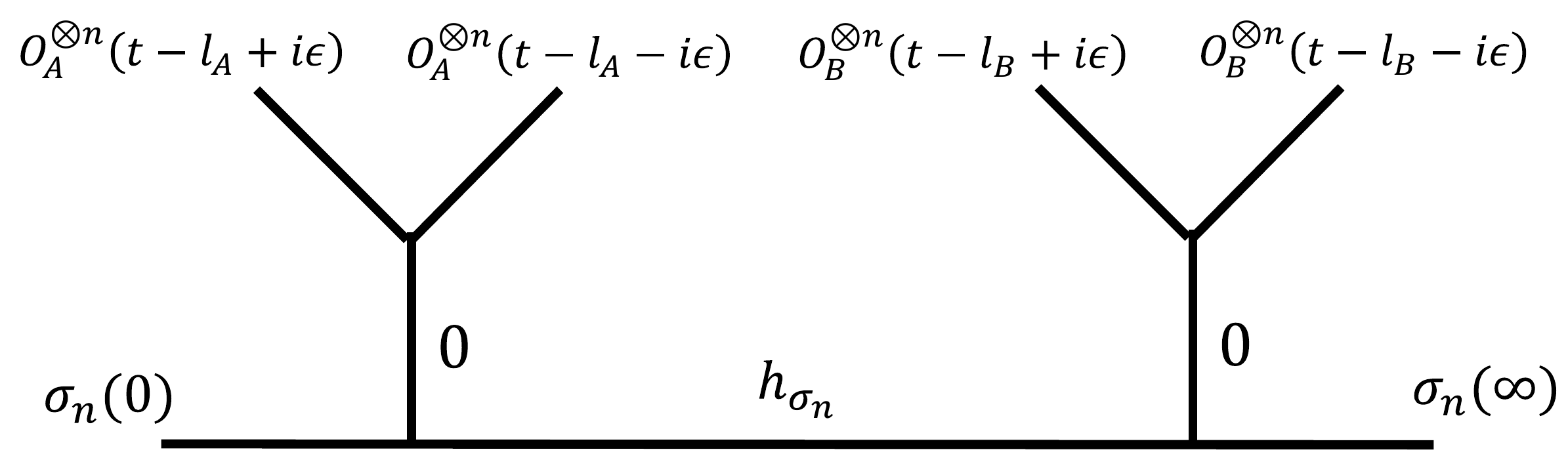}}
\newlength{\paw}
\settowidth{\paw}{\usebox{\boxpa}} 

\newsavebox{\boxpb}
\sbox{\boxpb}{\includegraphics[width=100pt]{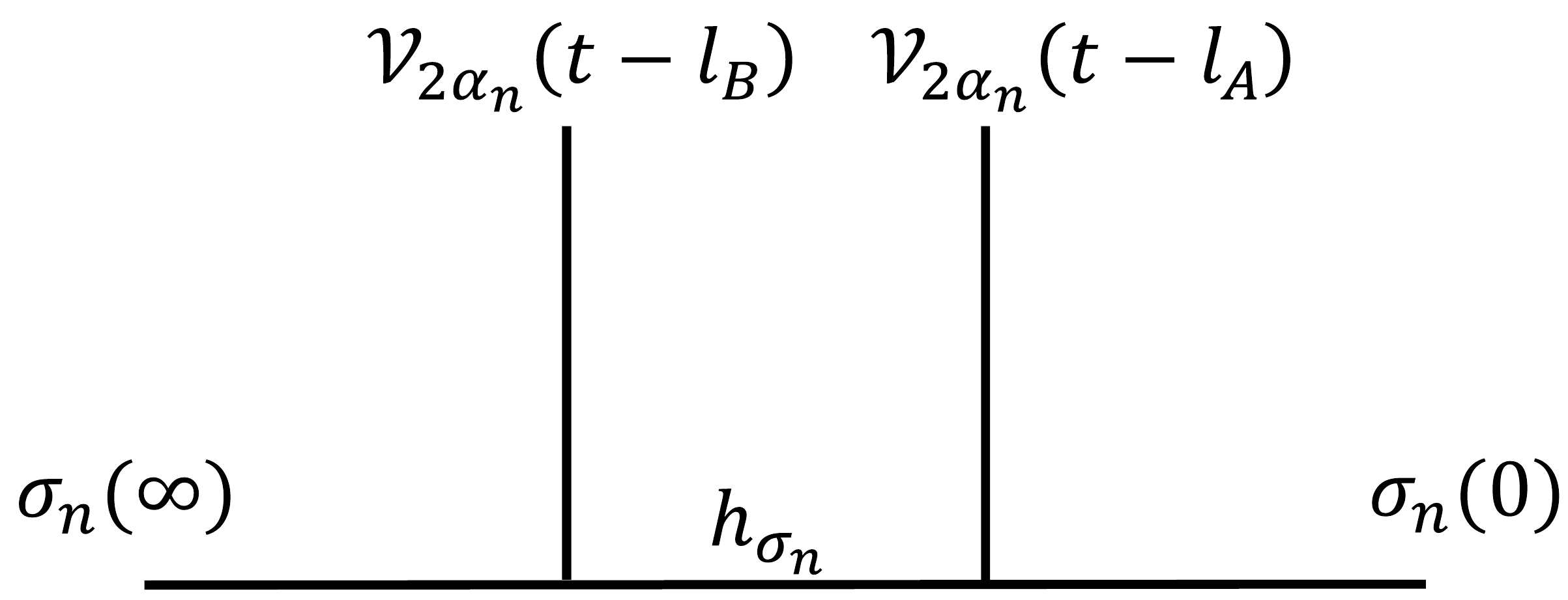}}
\newlength{\pbw}
\settowidth{\pbw}{\usebox{\boxpb}} 

\newsavebox{\boxpc}
\sbox{\boxpc}{\includegraphics[width=100pt]{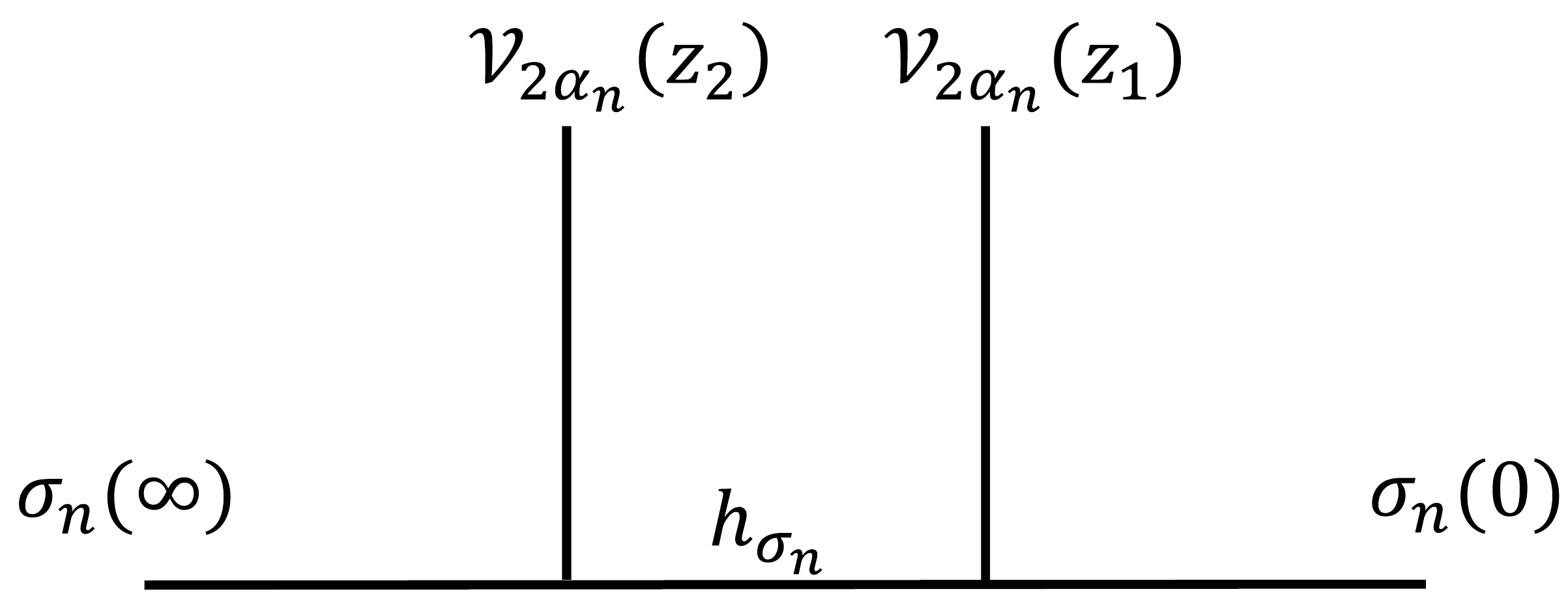}}
\newlength{\pcw}
\settowidth{\pcw}{\usebox{\boxpc}} 

\begin{equation}
\begin{aligned}
&\frac{\ave{O_A^{\otimes n}O_A^{\otimes n} O_B^{\otimes n}O_B^{\otimes n}   \sigma_n \bar{\sigma_n} }}{\ave{O_A^{\otimes n}O_A^{\otimes n} O_B^{\otimes n}O_B^{\otimes n} }\ave{\sigma_n \bar{\sigma_n}}}\\
& \ar{\e \to 0}
	 (2 i \ep)^{2h_{2\a_n}}  \parbox{\pbw}{\usebox{\boxpb}}\\
& \times\text{Res}\pa{  -2\pi i \   {{\bold M}^{(n)}}_{0, \a}[O_A];2\a_n }  
	\text{Res}\pa{  -2\pi i \   {{\bold M}^{(n)}}_{0, \a}[O_B];2\a_n },
\end{aligned}
\end{equation}
where we used the monodromy transformation (\ref{eq:monotrans}).
The remaining conformal block can not be evaluated for general $n$, however, if we restrict ourselves on the limit $n \to 1$,
this block has a simple form as
\begin{equation}
 \parbox{\pcw}{\usebox{\boxpc}} \ar{n-1 \to 0} \br{z_1\pa{z_2-z_1}}^{-2h_{\sigma_n}} .
\end{equation}
As a result we obtain
\begin{equation}
\begin{aligned}
& \frac{\ave{O_A^{\otimes n}O_A^{\otimes n} O_B^{\otimes n}O_B^{\otimes n}   \sigma_n \bar{\sigma_n} }}{\ave{O_A^{\otimes n}O_A^{\otimes n} O_B^{\otimes n}O_B^{\otimes n} }\ave{\sigma_n \bar{\sigma_n}}}\\
&\ar{\substack{\e \to 0 \\ \text{and} \\  n-1 \to 0}} \BR{\pa{\fr{2 i \e}{t-l_A}} \pa{\fr{2 i \e}{t-l_B}}}^{2h_{\a_n}}   \pa{\fr{l_B-l_A}{t-l_A}}^{-2h_{\sigma_n}}\\
	&\times 
	\text{Res}\pa{  -2\pi i \   {{\bold M}^{(n)}}_{0, \a}[O_A];2\a_n }  
	\text{Res}\pa{  -2\pi i \   {{\bold M}^{(n)}}_{0, \a}[O_B];2\a_n }.
\end{aligned}
\end{equation}
This result leads to the entanglement entropy for the double-excited state in the late time ($t>l_B$) as
\begin{equation}\label{eq:SAB}
\Delta S_A[O_A;O_B]=\Delta S_A[O_A] + \Delta S_A[O_B]+\fr{c}{6} \log \fr{l_B-l_A}{t-l_A}.
\end{equation}
Note that we can remove the restriction that the interval $A$ is large. If we choose the region $A$ to be the interval $[ 0, L]$ with $L>l_B$, then the result (\ref{eq:SAB}) when $L+l_B>t>l_B $ is replaced by
\begin{equation}\label{eq:SAB2}
\begin{aligned}
&\Delta S_A[O_A;O_B]\\
&=\Delta S_A[O_A] + \Delta S_A[O_B]+\fr{c}{6} \log \fr{(l_B-l_A)L}{(t-l_A)\pa{L-(t-l_B)}}.
\end{aligned}
\end{equation}
And the generalization to put the CFT on a circle with length $L_{\text{tot}}$ is 
\begin{equation}
\begin{aligned}
&\Delta S_A[O_A;O_B]\\
&=\Delta S_A[O_A] + \Delta S_A[O_B]+\fr{c}{6} \log \fr{\sin\pa{\pi\fr{l_B-l_A}{L_{\text{tot}}}} \sin\pa{\pi\fr{L}{L_{\text{tot}}}}}{ \sin\pa{\pi\fr{t-l_A}{L_{\text{tot}}}}  \sin\pa{\pi\fr{L-(t-l_B)}{L_{\text{tot}}}}},
\end{aligned}
\end{equation}
which can be obtained by the method developed in \cite{Calabrese2004}.

This is our main result in this paper. The interesting point is that this entanglement entropy (\ref{eq:SAB}) cannot be given by just a simple sum of the entanglement entropies for the single-excitation with $O_A$ and $O_B$. In particular, the time dependence in the late time limit ($t \gg \e, l_A, l_B$) is 
\begin{equation}
\begin{aligned}
\Delta S_A[O_A;O_B] &\sim \fr{c}{6} \log t,
\end{aligned}
\end{equation}
which is contrary to the naive expectation that each single-excitation contributes to the entanglement entropy as
\begin{equation}
\Delta S_A[O_A;O_B] \sim 2 \times \fr{c}{6} \log t.
\end{equation}
We expect that the appearance of the extra term besides $\Delta S_A[O_A]$ and $\Delta S_A[O_B]$ is the characteristic of the holographic CFT, and this term can be interpreted as an interaction part of the entanglement entropy. In other words, this part measures the interaction between two particles. If so, this interaction part is expected to be negative because this interaction should be dual to gravitational force between two particles, which is attractive (see \cite{Caputa2019}). 
Since $t>l_B>l_A$, our interaction part is negative, therefore, our result is perfectly consistent with this expectation.
Note that our result does not rely on the large $c$ assumption, in other words, our result is exact in central charge.

We emphasize that the formula (\ref{eq:SAB}) holds only in pure CFTs.
The point is that in pure CFTs, the dominant contribution of the integral over $\a$ in (\ref{eq:twistRenyi}) comes from $\a=2\a_n$,
 however, this is not the case in other CFTs. In RCFTs, the integral (\ref{eq:twistRenyi}) is dominated by the vacuum contribution $\a =0$, which leads to
\begin{equation}
\begin{aligned}
&\frac{\ave{O_A^{\otimes n}O_A^{\otimes n} O_B^{\otimes n}O_B^{\otimes n}   \sigma_n \bar{\sigma_n} }}{\ave{O_A^{\otimes n}O_A^{\otimes n} O_B^{\otimes n}O_B^{\otimes n} }\ave{\sigma_n \bar{\sigma_n}}} \ar{ \e \to 0} 
	 {{\bold M}^{(J,n)}}_{0, 0}[O_A]   {{\bold M}^{(J,n)}}_{0, 0}[O_B],
\end{aligned}
\end{equation}
where ${\bold M}^{(J,n)}$ is the monodromy matrix associated to $(J; \text{current algebra of seed RCFT})^n/\bb{Z}_n$.
In fact, this means that the entanglement entropy after the double-excitation is given by the sum rule,
\begin{equation}
\Delta S_A[O_A;O_B]=\Delta S_A[O_A] + \Delta S_A[O_B].
\end{equation}
This statement has already shown in \cite{Caputa2016a, Numasawa2016, Guo2018} in another way. For the same reason, this sum rule also appears in free massless scalar field theories  \cite{Nozaki2014a}. From these observations, we can expect that the appearance of the negative interaction term in the entanglement entropy implies the existence of Einstein gravity dual. 

\section{Discussion}

In this paper, we studied time evolution of entanglement entropy after the double local operator excitations. We assumed the absence of chiral primaries, which allows us to restrict ourselves to the vacuum exchange. We evaluated the deviation of entanglement entropy of double local excitations, from the sum of entanglement entropies of each single local excitation. The deviation was found to be negative, which can be interpreted as the attractive force in dual gravity, as was observed in \cite{Caputa2019} for joining and splitting quenches.

We emphasize that our CFT result is quite general, that is, we assumed neither large $c$ limit nor light excitations, but only pure CFT. And the result shows that the correction to the entanglement entropy from the gravitational interaction always exist in double local excitation set up, and the correction does not depend on how heavy these operators are.

We address possible gravity dual of the double locally quenched state. Let us consider the merger of two particles as a candidate, and compute the entanglement entropy. We put two conical defects at the boundary of Poincare AdS, with same mass and momentum perpendicular to the boundary. Such geometries in global AdS were constructed in \cite{Matschull:1998rv,Ageev2016,Arefeva2017}. Then these particles eventually form a black hole with horizon radius $r_h$, if they have large enough energy. We consider a subregion $[0, L]$ and the corresponding Ryu-Takayanagi surface. At late time, the Ryu-Takayanagi surface probes the region outside the horizon of the black hole, so the entanglement entropy for the subregion $[0,~L]$ at late time $l_A< l_B \ll t \ll L$ is given by
\begin{equation}\label{eq:EEmerger}
S_A[O_A;O_B]\underset{l_A< l_B \ll t \ll L}{\rightarrow} \frac{c}{3}\text{log}\frac{L}{\e}+\frac{c}{6}\text{log}\frac{t}{l_B-l_A}+\frac{c}{6}\text{log}\pa{\frac{\text{sinh}(\pi r_h)}{r_h}},\end{equation} 
where $r_h$ is the horizon radius of the black hole. Here we started with BTZ geometry
\begin{equation}
ds^2=-(r^2-r_h^2)d\tau^2+\frac{dr^2}{r^2-r_h^2}+r^2d\phi^2,
\end{equation}
with identification of two surfaces, which are given by
\begin{equation}
\sqrt{1-\frac{r_h^2}{r^2}}\text{cosh}\pa{r_h t}=\text{cosh}\pa{r_h \phi}\pm\text{tanh}\frac{\pi r_h}{2}\text{sinh}\pa{r_h t},
\end{equation}
for massless initial particles. The spacetime region surrounded by these two surfaces is now removed from the spacetime after identification. The intersections of these surfaces at $\phi=0, ~\pi$ correspond to the conical defects. We boosted this geometry so that the spacetime has Minkowski boundary, via coordinate transformation
\begin{equation}
\begin{aligned}\left\{
    \begin{array}{ll}
    \sqrt{1+r^2}\cos{\tau}=\frac{\text{e}^{\beta}+\text{e}^{-\beta}(z^2+x^2-t^2)}{2z}  ,\\
     \sqrt{1+r^2}\text{sin}\tau=\frac{t}{z},\\
     r\sin(\phi-\frac{\pi}{2})=\frac{x}{z},\\
     r\text{cos}(\phi-\frac{\pi}{2})=-\frac{-\text{e}^{\beta}+\text{e}^{-\beta}(z^2+x^2-t^2)}{2z},     \end{array}
  \right.
  \end{aligned}\end{equation}
  where $e^{\beta}=\frac{l_B-l_A}{2}$. 
  It is interesting to see whether the black hole merger described above or its variants, can reproduce the entanglement entropy (\ref{eq:SAB}) of double local quench. Indeed, both (\ref{eq:SAB}) and (\ref{eq:EEmerger}) grow in the late time as 
  \begin{equation}
  S_A[O_A;O_B] \sim  \frac{c}{6}\text{log}~t.
  \end{equation} This behavior can be naturally understood by noting two particle system in the bulk with gravitation, should look like one particle system at late time. On the other hand, the remaining terms which are constant in time, highly depend on the dynamics in the bulk, and the merger does not seem to reproduce (\ref{eq:SAB}). Another candidate of gravity dual is given by two massive particles moving in parallel, without merger which creates a new massive object. It is very interesting to explore the gravity dual of double local quench more quantitively, and find the dual which can reproduce (\ref{eq:SAB}) correctly.
  
Finally, we discuss future directions. It would be interesting to consider more refined tools of analysis than the single interval entanglement entropy. For example, the reflected entropy \cite{Dutta2019e} is recently calculated in a single local quench state \cite{Kusuki2019a}. This analysis can be generalized to a double local quench state, which would provide more information about the connection between CFT and gravity. It would be also interesting to consider negativity  \cite{Kudler-Flam2019b,Kusuki2019} and odd entanglement entropy \cite{Tamaoka2019}.


\section*{Acknowledgments}
We would like to thank Pawel Caputa, Tadashi Takayanagi and Kotaro Tamaoka for fruitful discussions and comments.
YK and MM are supported by the JSPS fellowship. 
We are grateful to the conference “Quantum Information and String Theory 2019” in YITP.

\bibliographystyle{JHEP}
\bibliography{renyi2}

\end{document}